\documentclass[aps,prl,twocolumn,superscriptaddress,preprintnumbers,%
               showpacs,nofootinbib]{revtex4}
\newcommand{\PRE}[1]{}       

\usepackage{bm}
\usepackage{epsfig}

\newcommand{\postscript}[2]{\setlength{\epsfxsize}{#2\hsize}
   \centerline{\epsfbox{#1}}}

\newcommand{\mev}{\text{MeV}}
\newcommand{\gev}{\text{GeV}}
\newcommand{\tev}{\text{TeV}}
\newcommand{\pb}{\text{pb}}

\newcommand{\cm}{\text{cm}}

\newcommand{\s}{\text{s}}

\newcommand{\sr}{\text{sr}}
\newcommand{\etal}{{\em et al.}}

\newcommand{\eqref}[1]{Eq.~(\ref{#1})}
\newcommand{\eqsref}[2]{Eqs.~(\ref{#1}) and (\ref{#2})}
\newcommand{\sla}[1]{\not{\! #1}}

\newcommand{\mB}{m_{B^1}}
\newcommand{\mq}{m_{q^1}}

\newcommand{\bold}[1]{{\text{\normalsize\bm{$#1$}}}}

\begin{document}

\preprint{EFI-02-95,UCI-TR-2002-23,UFIFT-HEP-02-21,CERN-TH/2002-157}

\title{
\PRE{\vspace*{1.5in}}
Kaluza-Klein Dark Matter
\PRE{\vspace*{0.3in}}
}

\author{Hsin-Chia Cheng}
\affiliation{Enrico Fermi Institute,
The University of Chicago, Chicago, IL 60637, USA
\PRE{\vspace*{.1in}}
}
\author{Jonathan L.~Feng}
\affiliation{Department of Physics and Astronomy,
University of California, Irvine, CA 92697, USA
\PRE{\vspace*{.1in}}
}

\author{Konstantin T.~Matchev%
\PRE{\vspace*{.2in}}
}
\affiliation{Department of Physics, University of Florida, Gainesville,
FL 32611, USA 
\PRE{\vspace*{.1in}}
}
\affiliation{Theory Division, CERN, CH-1211 Geneva, Switzerland
\PRE{\vspace*{.1in}}
}


\begin{abstract}
\PRE{\vspace*{.1in}} 
We propose that cold dark matter is made of Kaluza-Klein particles and
explore avenues for its detection.  The lightest Kaluza-Klein state is
an excellent dark matter candidate if standard model particles
propagate in extra dimensions and Kaluza-Klein parity is conserved.
We consider Kaluza-Klein gauge bosons.  In sharp contrast to the case
of supersymmetric dark matter, these annihilate to hard positrons,
neutrinos and photons with unsuppressed rates.  Direct detection
signals are also promising.  These conclusions are generic to bosonic
dark matter candidates.
\end{abstract}


\maketitle


The identity of dark matter is currently among the most profound
mysteries in particle physics, astrophysics, and cosmology.  Recent
data from supernovae luminosities, cosmic microwave anisotropies, and
galactic rotation curves all point
consistently to the existence of dark matter with mass density $\Omega
\approx 0.3$ relative to the critical density.  At the same time, all
known particles are excluded as dark matter candidates, making the
dark matter problem the most pressing phenomenological motivation for
particles and interactions beyond the standard model.

Among the myriad options, the possibility of particle dark matter with
weak interactions and weak-scale mass is particularly tantalizing.
Puzzles concerning electroweak symmetry breaking suggest that such
particles exist, and, if stable, their thermal relic density is
generically in the desired range.  Among these candidates, neutralinos
in supersymmetric theories are by far the most widely studied.
Neutralinos have spin 1/2 and are their own anti-particles; that is,
they are Majorana fermions.  They may be detected directly through
scattering in detectors, or indirectly through the decay products that
result when neutralinos annihilate in pairs. For indirect detection,
however, the Majorana nature of neutralinos implies that annihilation
is chirality-suppressed, leading to soft secondary positrons, photons,
and neutrinos, and considerably diminishing prospects for discovery.

Here we study a specific example of a generic alternative:~bosonic
cold dark matter.  If particles propagate in extra spacetime
dimensions, they will have an infinite tower of partner states with
identical quantum numbers, as noted long ago by Kaluza and
Klein~\cite{Kaluza:tu}.  We consider the case of universal extra
dimensions (UED)~\cite{Appelquist:2000nn}, in which all standard model
particles propagate.  Such models provide, in the form of stable
Kaluza-Klein (KK) partners, the only specific dark matter candidate to
emerge from theories with extra
dimensions~\cite{Dienes:1998vg,Cheng:2002iz,Cheng:2002ab}.  KK dark matter
generically has the desired relic
density~\cite{Servant:2002aq,Kolb:fm}.  Here we explore for the first
time the prospects for its detection.


For concreteness, we consider the simplest UED model, with one extra
dimension of size $R\sim\tev^{-1}$ compactified on an $S^1/Z_2$
orbifold.  At the lowest order, the KK masses are simply the momenta
along the extra dimension and are quantized in units of $1/R$. The
degeneracy at each KK level is lifted by radiative corrections and
boundary terms~\cite{Cheng:2002iz}.  The boundaries also break
momentum conservation in the extra dimension down to a $Z_2$ parity,
under which KK modes with odd KK number are charged.  This KK-parity
corresponds to the symmetry of reflection about the midpoint in the
extra dimension; it is anomaly-free and not violated by quantum
gravity effects.  KK-parity conservation implies that the lightest KK
particle is stable. KK partners of electroweak gauge bosons and
neutrinos are then all possible dark matter candidates. We consider
$B^1$, the first KK mode of the hypercharge gauge boson, which at
one-loop is naturally the lightest KK mass eigenstate in minimal
models~\cite{Cheng:2002iz,Cheng:2002ab}.

In this UED scenario, constraints from precision data require only
$1/R \agt 300~\gev$~\cite{Appelquist:2000nn}.  Collider searches are
also quite challenging: the Tevatron Run II may probe slightly beyond
this bound and the LHC may reach $1/R \sim
1.5~\tev$~\cite{Cheng:2002ab}.  Dark matter searches provide another
possibility for probing these models and differentiating them from
other new physics.


For a given KK spectrum, the $B^1$ thermal relic density may be
accurately determined~\cite{Servant:2002aq}.  $B^1$s annihilate
effectively through $s$-wave processes, unlike neutralinos, and so the
desired thermal relic density is obtained for higher masses than
typical for neutralinos.  If $B^1$s are the only KK modes with
significant abundance at the freeze-out temperature, the desired relic
density is found for $\mB \approx 1~\tev$.  However, many other KK
states may be closely degenerate with $B^1$, and their presence at
freeze-out will modify this conclusion.  KK quarks and gluons
annihilate with much larger cross sections through strong
interactions, and so increase the predicted $\mB$.  On the other hand,
degenerate KK leptons lower the average annihilation cross section and
require lower $\mB$. In addition to the cosmological assumptions
present in all relic density calculations, the $B^1$ relic density is
therefore rather model-dependent, with the optimal $\mB$ ranging from
several hundred GeV to a few TeV, depending sensitively on the KK
spectrum.  Here we study the prospects for detection in a
model-independent way by considering $\mB$ as a free parameter in the
appropriate range.


We first consider the direct detection of $B^1$ dark matter.  Dark
matter particles are currently non-relativistic, with velocity $v \sim
10^{-3}$.  For weak-scale dark matter, the recoil energy from
scattering off nuclei is $\sim 0.1~\mev$, and far less for scattering
off electrons.  We therefore consider elastic scattering off nucleons
and nuclei.

At the quark level, $B^1$ scattering takes place through KK quarks,
with amplitude ${\cal M}_q^{q^1} = {\cal M}_{q_L}^{q^1} + {\cal
M}_{q_R}^{q^1}$, where
\begin{eqnarray}
\lefteqn{{\cal M}_{q_i}^{q^1} = - i \frac{e^2}{\cos^2 \theta_W} Y_{q_i}^2
\varepsilon_{\mu}^{\ast}(p_3) \varepsilon_{\nu}(p_1) \times }
\nonumber \\
&& \bar{u} (p_4) \! \left[ 
\frac{\gamma^{\mu} \! \sla{k}_1 \gamma^{\nu}} {k_1^2 - m_{q_i^1}^2} 
+ \frac{\gamma^{\nu} \! \sla{k}_2 \gamma^{\mu}} {k_2^2 - m_{q_i^1}^2} 
\right] \! P_i \, u(p_2) \ , 
\end{eqnarray}
$Y=Q-I$ is hypercharge, $k_1 = p_1 + p_2$, and $k_2 = p_2 - p_3$; and
through Higgs exchange, with amplitude
\begin{equation}
{\cal M}^h_q = i \frac{e^2}{2 \cos^2 \theta_W} 
\frac{m_f}{k_3^2 - m_h^2} \varepsilon_{\mu}^{\ast}(p_3) 
\varepsilon^{\mu}(p_1) \bar{u}(p_4) u(p_2) \ ,
\end{equation}
where $k_3 = p_1 - p_3$.  In the extreme non-relativistic limit, $p_1
= p_3 = (\mB, \bold{0})$, and expanding to linear order in $p_2 =
(E_q, \bold{p}_q)$, these amplitudes then reduce to
\begin{eqnarray}
{\cal M}_{q}^{q^1}
&\approx& \alpha_{q}
\varepsilon_{\mu}^\ast(p_3) \varepsilon_{\nu}(p_1)
\varepsilon^{0\mu \nu \rho} 
\xi_4^{\dagger} \frac{\sigma_{\rho}}{2} \xi_2 \nonumber \\
&& - i \beta_q \varepsilon_{\mu}^{\ast}(p_3) 
\varepsilon^{\mu}(p_1) \xi_4^\dagger \xi_2  \\
{\cal M}_{q}^{h} &\approx&
 - i \gamma_q \varepsilon_{\mu}^{\ast}(p_3) 
\varepsilon^{\mu}(p_1) \xi_4^\dagger \xi_2 \ ,
\end{eqnarray}
where $\xi_4$ and $\xi_2$ are two-component spinors, and
\begin{eqnarray} 
\alpha_q \! \! &=& \! \! \frac{2 e^2}{\cos^2 \theta_W} \left[ 
\frac{Y_{q_L}^2 \mB}{m_{q_L^1}^2 - \mB^2} +
(L \to R) \right]  \label{alpha} \\
\beta_q \! \! &=& \! \! E_q \frac{e^2}{\cos^2 \theta_W} 
\left[ Y_{q_L}^2 \frac{\mB^2 + m_{q_L^1}^2}{(m_{q_L^1}^2 - \mB^2)^2} 
+ (L \to R) \right] 
\label{beta} \\
\gamma_q \! \! &=& \! \! m_q \frac{e^2}{2 \cos^2 \theta_W}
\frac{1}{m_h^2} \ .
\end{eqnarray}
The interactions divide into spin-dependent and spin-independent
parts~\cite{Goodman:1984dc}. Higgs exchange contributes to scalar
couplings, while $q^1$ exchange contributes to both.  Note that the
two contributions to scalar interactions interfere constructively;
barring extremely heavy KK masses, there is an inescapable lower bound
on both spin-dependent and scalar cross sections.

The spin-dependent coupling is $\alpha_{q} \bold{S}_{B^1} \cdot
\bold{S}_{q}$, where $\bold{S}_{B^1}$ and $\bold{S}_{q}$ are spin
operators.  We must evaluate this matrix element between nucleon or
nucleus bound states.  By the Wigner-Eckart theorem, we may replace
$\bold{S}_{q}$ by $\lambda_q \bold{J}_N$, where $\bold{J}_N$ is the
nucleon or nuclear spin operator.  The constant of proportionality is
\begin{equation}
\lambda_q = \Delta_q^p \langle S_p \rangle/J_N 
+ \Delta_q^n \langle S_n \rangle/J_N \ .
\label{lambda}
\end{equation}
$\Delta_q^{p,n}$ is given by $\langle p,n | \bold{S}^{\mu}_q | p, n
\rangle \equiv \Delta_q^{p,n} \bold{S}^{\mu}_{p,n}$ and is the
fraction of the nucleon spin carried by quark $q$. A recent analysis
finds $\Delta_u^p = \Delta_d^n = 0.78 \pm 0.02$, $\Delta_d^p =
\Delta_u^n = -0.48 \pm 0.02$, and $\Delta_s^p = \Delta_s^n = -0.15 \pm
0.02$~\cite{Mallot:1999qb}.  $\langle S_{p,n} \rangle / J_N \equiv
\langle N | S_{p,n} | N \rangle / J_N$ is the fraction of the total
nuclear spin $J_N$ that is carried by the spin of protons or neutrons.
For scattering off protons and neutrons, $\lambda_q$ reduces to
$\Delta_q^p$ and $\Delta_q^n$, respectively.

The spin-dependent cross section is $m_N^2/[4\pi (\mB + m_N)^2]
\langle | {\cal M}|^2 \rangle$, where ${\cal M} = \sum_q {\cal M}_q$
and $\langle \ \rangle$ denotes an average over initial polarizations
and sum over final polarizations.  Using $\langle (\bold{S}_{B^1}
\cdot \bold{J}_N)^2 \rangle = \frac{2}{3} J_N (J_N+1)$, we find
\begin{equation}
\sigma_{\text{spin}} = \frac{1}{6\pi} \frac{m_N^2}{(\mB + m_N)^2}
J_N (J_N+1) \bigg[ \sum_{u,d,s} \alpha_q \lambda_q \bigg]^2 \ ,
\label{sigma_spin}
\end{equation}
where $\alpha_q$ and $\lambda_q$ are given in \eqsref{alpha}{lambda}.

The spin-independent cross section is
\begin{equation}
\sigma_{\text{scalar}} =  \frac{m_N^2}{4\pi\, (\mB + m_N)^2} 
\left[Z f_p +(A-Z) f_n\right]^2 \ ,
\end{equation}
where $Z$ and $A$ are nuclear charge and atomic number,
\begin{equation}
f_p = \sum_{u, d, s} (\beta_q + \gamma_q) 
\langle p | \bar{q} q | p \rangle 
= \sum_{u, d, s} \frac{\beta_q + \gamma_q}{m_q} m_p 
f^p_{T_q} \ ,
\label{si}
\end{equation}
and similarly for $f_n$.  We take $f^{p}_{T_u}=0.020\pm 0.004$,
$f^{p}_{T_d}=0.026\pm 0.005$, $f^{n}_{T_u}=0.014\pm 0.003$,
$f^{n}_{T_d}=0.036\pm 0.008$, and $f^{p,n}_{T_s}=0.118\pm
0.062$~\cite{Ellis:2000ds}. $E_q$ of \eqref{beta} is the energy of a
bound quark and is rather ill-defined.  In evaluating \eqref{si}, we
conservatively replace $E_q$ by the current mass $m_q$.  We also
neglect couplings to gluons mediated by heavy quark loops; a detailed
loop-level analysis along the lines of
Refs.~\cite{Drees:1993bu,Drees:1992am} for neutralinos is in
progress~\cite{inprogress}.  Given the constructive interference noted
above, these contributions can only increase the cross section.

\begin{figure}[tbp]
\postscript{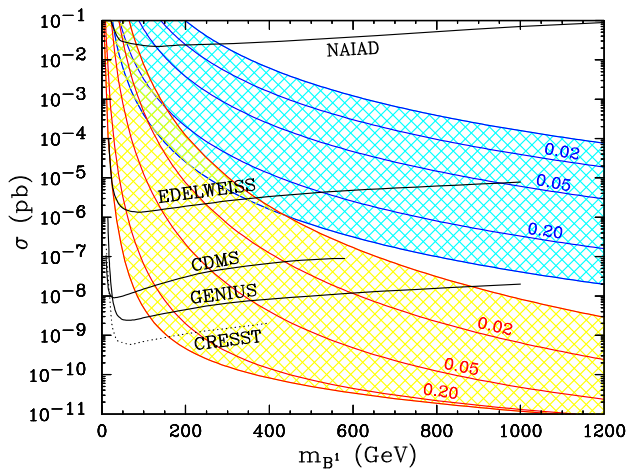}{0.95}
\caption{Predicted spin-dependent proton cross sections (dark-shaded, blue),
along with the projected sensitivity of a 100 kg NAIAD
array~\cite{Spooner:kt}; and predicted spin-independent proton cross
sections (light-shaded, red), along with the current EDELWEISS
sensitivity~\cite{Benoit:2002hf}, and projected sensitivities of
CDMS~\cite{Schnee:gf}, GENIUS~\cite{Klapdor-Kleingrothaus:2000eq}, and
CRESST~\cite{Bravin:1999fc}.  (The CRESST projection is long-term and
conditional upon increased exposure and improved background
rejection.) The predictions are for $m_h = 120~\gev$ and $0.01 \le r =
(\mq - \mB) / \mB \le 0.5$, with contours for specific intermediate 
$r$ labeled.
\label{fig:direct} }
\end{figure}

We present both spin-independent and spin-dependent cross sections in
Fig.~\ref{fig:direct}.  We assume that all first level KK quarks are
degenerate with mass $\mq$.  Proton cross sections are given; neutron
cross sections are similar for spin-dependent interactions and almost
identical for scalar cross sections.  The cross sections are large for
low $\mB$.  They are also strikingly enhanced by $r^{-2}$ for small $r
\equiv (\mq - \mB) / \mB$ when scattering takes place near an
$s$-channel pole.  Such degeneracy is unmotivated in general, but is
natural for UED models, where all KK particles are highly degenerate
at tree-level.

Projected sensitivities of near future experiments are also shown in
Fig.~\ref{fig:direct}.  For scattering off individual nucleons, scalar
cross sections are suppressed relative to spin-dependent ones by $\sim
m_p/\mB$.  However, this effect is compensated in large nuclei where
spin-independent rates are enhanced by $\sim A^2$.  In the case of
bosonic KK dark matter, the latter effect dominates, and the
spin-independent experiments have the best prospects for detection,
with sensitivity to $\mB$ far above current limits.

Dark matter may also be detected when it annihilates in the galactic
halo, leading to positron excesses in space-based and balloon
experiments.  The positron flux is~\cite{Moskalenko:1999sb}
\begin{equation}
\frac{d\Phi_{e^+}}{d\Omega dE} = \frac{\rho^2}{\mB^2}
\sum_i \langle \sigma_i v \rangle B_{e^+}^i
\! \! \int \! \! dE_0 f_i(E_0) G(E_0, E) \ ,
\label{dPhi}
\end{equation}
where $\rho$ is the local dark matter mass density, the sum is over
all annihilation channels $i$, and $B_{e^+}^i$ is the $e^+$ branching
fraction in channel $i$. The initial positron energy distribution is
given by $f(E_0)$, and the Green function $G(E_0, E)$ propagates
positrons in the galaxy.

Several channels contribute to the positron flux.  Here we focus on
the narrow peak of primary positrons from direct $B^1 B^1\rightarrow
e^+ e^-$ annihilation.  (Annihilation to muons, taus and heavy quarks
also yield positrons through cascade decays, but with relatively soft
and smeared spectra.)  In this case, the source is monoenergetic, and
\eqref{dPhi} simplifies to
\begin{eqnarray}
\lefteqn{\frac{d\Phi_{e^+}}{d\Omega dE} = 2.7\times 10^{-8} 
\cm^{-2} \s^{-1} \sr^{-1} \gev^{-1} 
\frac{\langle\sigma_{ee} v \rangle}{\pb} }
\nonumber \\
&&\times
\left[ \frac{\rho}{0.3~\gev/\cm^3}\right]^2
\left[ \frac{1~\tev}{\mB}\right]^2
g\left(1,\frac{E}{\mB}\right) ,
\end{eqnarray}
where the annihilation cross section is
\begin{equation}
\langle \sigma_{ee} v \rangle = \frac{e^4}{9\pi \cos^4 \theta_W}
\left[ \frac{Y_{e^1_L}^4}{\mB^2+m_{e^1_L}^2} 
+ (L \to R) \right]\ ,
\label{sigma_ee}
\end{equation}
and the reduced Green function $g$ is as in Ref.~\cite{Feng:2001zu}.

\begin{figure}[tbp]
\postscript{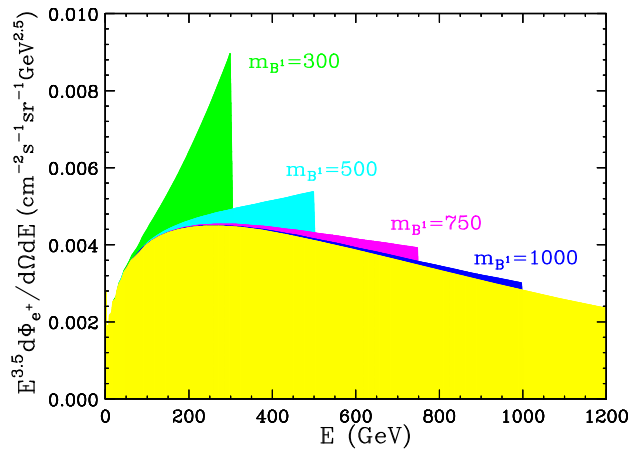}{0.95}
\caption{Predicted positron signals (dark shaded) above background
(light shaded) as a function of positron energy for $\mB = m_{e^1_L} =
m_{e^1_R} = 100$, 500, 750, and 1000 GeV.
\label{fig:positrons} }
\end{figure}

Positron spectra and an estimated background (model C {}from
Ref.~\cite{Moskalenko:1999sb}) are given in Fig.~\ref{fig:positrons}.
The sharp peak at $E_{e^+} = \mB$ is spectacular --- while propagation
broadens the spectrum, the mono-energetic source remains evident.
This feature is extremely valuable, as the background, although
resulting from many sources, should be smooth. Maximal $E_{e^+}$ also
enhances detectability since the background drops rapidly with energy.
Both of these virtues are absent for neutralinos, where Majorana-ness
implies helicity-suppressed annihilation amplitudes, and positrons are
produced only in cascades, leading to soft, smooth
spectra~\cite{Ellis:2001hv}.  A peak in the $e^+$ spectrum will not
only be a smoking gun for $B^1$ dark matter, it will also exclude
neutralinos as the source.

Of the many positron experiments, the most promising is
AMS~\cite{Barrau:2001ux}, the anti-matter detector to be placed on the
International Space Station.  AMS will distinguish positrons from
electrons even at 1 TeV energies~\cite{Hofer:1998sx}.  With aperture
$6500~\cm^2\sr$ and a runtime of 3 years, AMS will detect $\sim 1000$
positrons with energy above 500 GeV, and may detect a positron peak
from $B^1$ dark matter.


Photons from dark matter annihilation in the center of the galaxy also
provide an indirect signal.  The line signal from $B^1 B^1 \to \gamma
\gamma$ is loop-suppressed, and so we consider continuum photon
signals.  The integrated photon flux above some photon energy
threshold $E_{th}$ is~\cite{Feng:2001zu}
\begin{eqnarray}
\lefteqn{\Phi_{\gamma} (E_{th})=  5.6 \times 10^{-12}~\cm^{-2}~\s^{-1}
\bar{J}(\Delta \Omega) \, \Delta \Omega} \nonumber \\
&&\times 
\left[ \frac{1~\tev}{\mB} \right]^2
\sum_q \frac{\langle\sigma_{qq} v\rangle}{{\rm pb}}
\int_{E_{th}}^{\mB} 
\! \! dE \frac{dN_{\gamma}^q}{dE}\ ,
\label{phigamma}
\end{eqnarray}
where the sum is over all quark pair annihilation channels (with cross
sections similar to Eq.~(\ref{sigma_ee})), and $dN_{\gamma}^q/dE$ is
the differential gamma ray multiplicity for channel $qq$. The hardest
spectra result from fragmentation of light
quarks~\cite{Bergstrom:1997fj}, and so the lack of chirality
suppression again gives a relative enhancement over neutralinos.
$\Delta \Omega$ is the solid angle of the field of view of a given
telescope, and $\bar{J}$ is a measure of the cuspiness of the galactic
halo density profile.  There is a great deal of uncertainty in
$\bar{J}$, with possible values in the range 3 to $10^5$. We choose
$\Delta \Omega = 10^{-3}$ and a moderate value of $\bar{J} = 500$.

\begin{figure}[tbp]
\postscript{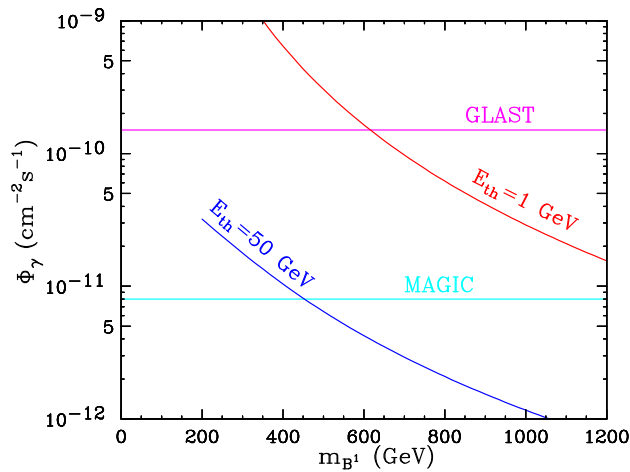}{0.95}
\caption{Integrated photon flux as a function of $\mB$ for energy
thresholds of 1 and 50 GeV.  Projected sensitivities for GLAST and
MAGIC are also shown.
\label{fig:photons} }
\end{figure}

Integrated photon fluxes are given in Fig.~\ref{fig:photons} for two
representative $E_{th}$:~1~GeV, accessible to space-based detectors,
and 50~GeV, characteristic of ground-based telescopes.  Estimated
sensitivities for two of the more promising experiments,
GLAST~\cite{Sadrozinski:wu} and MAGIC~\cite{MAGIC}, are also shown.
We find that photon excesses are detectable for $\mB \alt 600~\gev$.
Note that these signals may be greatly enhanced for clumpy halos with
large $\bar{J}$.


Finally, high-energy neutrinos from annihilating dark matter trapped
in the core of the Sun or the Earth, can be detected through their
charged-current conversion to muons. Unlike the case in supersymmetry,
$B^1$s can annihilate directly to neutrinos, with branching ratio
$\approx 1.2\%$. Secondary neutrinos may also result from final states
with heavy quarks, charged leptons, or Higgs bosons. Considering
primary neutrinos and those from tau decays from the Sun (which is
typically full, with capture and annihilation in equilibrium), we find
that, for $r = 0.5\, (0.02)$, next generation neutrino telescopes like
AMANDA, NESTOR and ANTARES will probe $\mB$ up to 200 GeV (600 GeV)
and IceCube will be sensitive to $\mB = 400$ GeV (1400
GeV)~\cite{inprogress}.


In conclusion, we find excellent prospects for KK dark matter
detection.  The elastic scattering cross sections are enhanced near
$s$-channel KK resonances, providing good chances for {\em direct}
detection.  In addition, {\em indirect} detection is typically much
more promising than in supersymmetry for three reasons. First, there
is no helicity suppression for the annihilation of bosonic KK dark
matter into fermion pairs.  Second, the preferred $B^1$ mass range is
higher than in supersymmetry, leading to harder positron, photon, and
neutrino spectra, with better signal-to-background ratio. And third,
$B^1$ annihilation produces primary positrons and neutrinos with
distinctive energy spectrum shapes, again facilitating observation
above background. Kaluza-Klein gauge bosons therefore provide a
promising and qualitatively new possibility for dark matter and dark
matter searches.



\begin{thebibliography}{99}

\bibitem{Kaluza:tu} T.~Kaluza,
Sitzungsber.\ Preuss.\ Akad.\ Wiss.\ Berlin (Math.\ Phys.\ ) 
{\bf K1}, 966 (1921);
O.~Klein,
Z.\ Phys.\  {\bf 37}, 895 (1926)
[Surveys High Energ.\ Phys.\  {\bf 5}, 241 (1986)].

\bibitem{Appelquist:2000nn}
T.~Appelquist, H.-C.~Cheng and B.~A.~Dobrescu,
Phys.\ Rev.\ D {\bf 64}, 035002 (2001)
[hep-ph/0012100].

\bibitem{Dienes:1998vg}
K.~R.~Dienes, E.~Dudas and T.~Gherghetta,
Nucl.\ Phys.\ B {\bf 537}, 47 (1999)
[hep-ph/9806292].

\bibitem{Cheng:2002iz}
H.-C.~Cheng, K.~T.~Matchev and M.~Schmaltz,
Phys.\ Rev.\ D {\bf 66}, 036005 (2002)
[hep-ph/0204342].

\bibitem{Cheng:2002ab}
H.-C.~Cheng, K.~T.~Matchev and M.~Schmaltz,
hep-ph/0205314.

\bibitem{Servant:2002aq}
G.~Servant and T.~M.~Tait,
hep-ph/0206071.

\bibitem{Kolb:fm}
See also E.~W.~Kolb and R.~Slansky,
Phys.\ Lett.\ B {\bf 135}, 378 (1984);
J.~Saito,
Prog.\ Theor.\ Phys.\  {\bf 77}, 322 (1987).

\bibitem{Goodman:1984dc}
M.~W.~Goodman and E.~Witten,
Phys.\ Rev.\ D {\bf 31}, 3059 (1985).

\bibitem{Mallot:1999qb}
G.~K.~Mallot,
Int.\ J.\ Mod.\ Phys.\ A {\bf 15S1}, 521 (2000).

\bibitem{Ellis:2000ds}
J.~R.~Ellis, A.~Ferstl and K.~A.~Olive,
Phys.\ Lett.\ B {\bf 481}, 304 (2000)
[hep-ph/0001005].


\bibitem{Drees:1993bu}
M.~Drees and M.~Nojiri,
Phys.\ Rev.\ D {\bf 48}, 3483 (1993).

\bibitem{Drees:1992am}
M.~Drees and M.~M.~Nojiri,
Phys.\ Rev.\ D {\bf 47}, 376 (1993).

\bibitem{inprogress}
H.-C.~Cheng, J.~L.~Feng and K.~T.~Matchev,
in progress.

\bibitem{Spooner:kt}
N.~J.~Spooner {\it et al.},
Phys.\ Lett.\ B {\bf 473}, 330 (2000).

\bibitem{Benoit:2002hf}
A.~Benoit {\it et al.},
astro-ph/0206271.

\bibitem{Schnee:gf}
R.~W.~Schnee {\it et al.},
Phys.\ Rept.\  {\bf 307}, 283 (1998).

\bibitem{Klapdor-Kleingrothaus:2000eq}
H.~V.~Klapdor-Kleingrothaus,
hep-ph/0104028.

\bibitem{Bravin:1999fc}
M.~Bravin {\it et al.}  [CRESST-Collaboration],
Astropart.\ Phys.\  {\bf 12}, 107 (1999)
[hep-ex/9904005].

\bibitem{Moskalenko:1999sb}
I.~V.~Moskalenko and A.~W.~Strong,
Phys.\ Rev.\ D {\bf 60}, 063003 (1999)
[astro-ph/9905283].


\bibitem{Feng:2001zu}
J.~L.~Feng, K.~T.~Matchev and F.~Wilczek,
Phys.\ Rev.\ D {\bf 63}, 045024 (2001)
[astro-ph/0008115].

\bibitem{Ellis:2001hv}
J.~R.~Ellis \etal,
Eur.\ Phys.\ J.\ C {\bf 24}, 311 (2002).

\bibitem{Barrau:2001ux}
A.~Barrau  [AMS Collaboration],
astro-ph/0103493.

\bibitem{Hofer:1998sx}
H.~Hofer and M.~Pohl,
Nucl.\ Instrum.\ Meth.\ A {\bf 416}, 59 (1998)
[hep-ex/9804016].

\bibitem{Bergstrom:1997fj}
L.~Bergstrom, P.~Ullio and J.~H.~Buckley,
Astropart.\ Phys.\  {\bf 9}, 137 (1998)
[astro-ph/9712318].

\bibitem{Sadrozinski:wu}
H.~F.~Sadrozinski,
Nucl.\ Instrum.\ Meth.\ A {\bf 466}, 292 (2001).

\bibitem{MAGIC}
MAGIC Collaboration, M.~Martinez {\em et al.}, OG.4.3.08 in {\em
Proceedings of ICRC99}, Utah, 17-25 August 1999.

\end{thebibliography}
\end{document}